\documentclass[pra,twocolumn,showpacs,10pt]{revtex4}%
\usepackage{amsfonts}
\usepackage{amsmath}
\usepackage{amssymb}
\usepackage{graphicx}%
\begin{document}
\title[Entanglement Manipulation between Quantum Spin Systems]{Entanglement manipulation via dynamics in multiple quantum spin systems}
\author{Andrea Casaccino}
\affiliation{Computer Architecture Group Lab, Information Engineering
Department, University of Siena, 53100 Siena, Italy\\
Research Laboratory of Electronics, Massachusetts Institute of Technology,
Cambridge, 020139, USA}
\author{Stefano Mancini}
\affiliation{School of Science and Technology, University of Camerino, 62032 Camerino, Italy\\
and INFN, Sezione di Perugia, 06123 Perugia, Italy}
\author{Simone Severini}
\affiliation{Department of Physics and Astronomy, University College London, WC1E 6BT
London, United Kingdom}

\begin{abstract}
We study manipulation of entanglement between two identical networks of
quantum mechanical particles. Firstly, we reduce the problem of entanglement
transfer to the problem of quantum state transfer. Then, we consider
entanglement concentration and purification based on free dynamics on the
networks and local measurements on the vertices. By introducing an appropriate
measure of efficiency, we characterize the performance of the protocol. We give
evidence that such a measure does not depend on the network topology, and we
estimate the contribution given by the number of entangled pairs initially shared
by the two networks.

\end{abstract}
\maketitle

\section{Introduction}

Networks of quantum mechanical particles play a fundamental role in extending
quantum information processing to multipartite settings and in foreseeing the
realization of distributed protocols for multi-users devices \cite{K08}.

In the present work we study manipulation of entanglement between quantum
networks and attempt to isolate useful graph-theoretic properties in relation
to the dynamics of entanglement. Specifically, we analyze the problem of
entanglement transfer through networks and then we relate its setting to the
fidelity problem associated with state transfer.

Our goal is to study entanglement concentration and purification, when the
evolution of the system involves two networks of spin-half particles with
$XY$-type interaction. Entanglement concentration (and, more generally,
purification) is among the most relevant tasks in quantum information
processing \cite{dur07}. It is a process extracting strongly entangled pairs
out of initially weakly entangled ones, only by mean of local operations and
classical communication. Most of the known protocols involve control-not
operations. In a similar way, multipartite measurements and classical
communication are used to propagate entanglement in certain type of quantum
networks \cite{mingetal, acin2}. Entanglement concentration has also been
obtained by exploiting the effects of quantum statistics of indistinguishable
particles \cite{omar}. A streaming (sequential) protocol for universal
entanglement concentration has recently appeared in \cite{got}, where a number
of input systems are processed in sequence, and perfect entangled pairs are
obtained at the end of the protocol.

Following \cite{Koji}, we simply exploit free dynamics on the networks and local
measurements on the sites. Somehow contrarily to a claim proposed in
\cite{Koji}, we show that this method works reasonably well also in the case of
$XY$-type interaction. We consider generic network topologies other than linear
spin chains and we introduce an efficiency measure to characterize the
performance of the protocol. It will turn out that, at least in the cases considered,
the structure of the networks only affects the time at which the maximum
efficiency is attained.

The topic of this paper fits into a growing literature. State and energy transfer in
spin systems are now areas of extensive study \cite{BB07} (see also \cite{PST}).
Protocols for creating entanglement between distant sites in a quantum network
have been designed in \cite{lape}, where the network considered is a lattices,
whose vertices share pure, non-maximally entangled pairs of qubits. The problem
of preparing maximally entangled states between different sites in a distributed
quantum system is related to classical percolation in statistical mechanics
\cite{acin}. In \cite{cuq}, the study of entanglement percolation is extended to
structures with precise statistical properties but unknown local topology. In
\cite{broad}, entanglement percolation is used for long-distance singlet
generation in networks with adjacent vertices connected by partially entangled
bipartite mixed states.

The structure of the paper is as follows. In Section II, we introduce the
considered networks model relying on the basic $XY$-type interaction. In
Section III, we give rigorous results for the problem of entanglement transfer
in quantum networks by considering the single excitation model. The quality of
the entanglement transmission is then related to the fidelity achieved between
the considered nodes, therefore extending the cases studied in
\cite{PST}. In Section IV, we discuss how to enhance the entanglement
of a pure states through free dynamics and local measurements on the nodes. We
introduce an efficiency measure for our protocols, by considering all possible
measurement outcomes and the corresponding probabilities. Finally, in Section
V, we show how the protocol works in the case of mixed entangled states by
considering Werner mixtures \cite{werner89}. Likewise Section IV, the
efficiency of the procedure will be related to the minimal number of initially
entangled pairs necessary to successfully complete the protocol. Conclusions
are drawn in Section VI, where we summarize the results and mention
potential physical realizations.


\section{Networks model}

Let $G=(V,E)$ be a simple undirected graph (that is, without loops or parallel
edges), with set of vertices $V(G)$ and set of edges $E(G)$. The
\emph{adjacency matrix} of $G$ is denoted by $\mathbf{\emph{A}}(G)$ and
defined by $[\mathbf{\emph{A}}(G)]_{ij}=1$, if $ij\in E(G)$; $[\emph{A}%
(G)]_{ij}=0$ if $ij\notin E(G)$. The adjacency matrix is a useful tool to describe a
network of $n$ spin-half quantum particles. The particles are usually associated
to the vertices of $G$, while the edges of $G$ represent their allowed couplings.
If one considers the $XY$ interaction model, then $\{i,j\}\in E(G)$ means that the
particles $i$ and $j$ interact by the Hamiltonian $[H(G)]_{ij}=-\left(
X_{i}X_{j}+Y_{i}Y_{j}\right)  $, where $X_{i}$, $Y_{i}$ are the Pauli operators of the
$i$-th particle (here we consider unit negative coupling constant). Thus, the
Hamiltonian of the whole network reads
\begin{equation}
H(G)=-\frac{1}{2}\sum_{i\neq j}[\mathbf{\emph{A}}(G)]_{ij}\left(  X_{i}%
X_{j}+Y_{i}Y_{j}\right)  \label{Hnet}%
\end{equation}
and it acts on the Hilbert space $\mathcal{H}=\left(  \mathbb{C}^{2}\right)
^{\otimes n}$.

In the following we will consider two networks with underlying graph $G_{A}$
and $G_{B}$ having, for the sake of simplicity, an equal number of nodes $n$.
Then, if we apply $XY$ model we have two Hamiltonians
\begin{align}
H(G_{A}) & =-\frac{1}{2}\sum_{i,j}^{n}[\mathbf{\emph{A}}(G_{A})]_{ij}(X_{A_{i}%
}X_{A_{j}}+Y_{A_{i}}Y_{A_{j}}),\nonumber\\
H(G_{B}) & =-\frac{1}{2}\sum_{i,j}^{n}[\mathbf{\emph{A}}(G_{B})]_{ij}(X_{B_{i}%
}X_{B_{j}}+Y_{B_{i}}Y_{B_{j}}).\label{HA&B}%
\end{align}


\section{Entanglement transfer}

Let $A$ and $B$ be two networks sharing a pair of qubits in a maximally
entangled state
\begin{equation}
\frac{1}{\sqrt{2}}(|0\rangle|1\rangle+|1\rangle|0\rangle).
\end{equation}
Suppose that the entangled qubits are in the \emph{i}-th site of the network
$A$ and in the \emph{k}-th site of the network $B$. We assume that all the
remaining qubits are in the ground state. Then the state for the joint system
$A+B$ will be
\begin{equation}
\frac{1}{\sqrt{2}}(|0\rangle_{A}|k\rangle_{B}+|i\rangle_{A}|0\rangle
_{B}),\label{2net}%
\end{equation}
where $|i\rangle_{A}$ denotes the state $|0\rangle_{1}|0\rangle_{2}%
\ldots|1\rangle_{i}\ldots|0\rangle_{n-1}|0\rangle_{n}$ for the network $A$; a
similar notation is used for the network $B$.

The evolution of the whole system will be governed by the Hamiltonians
\eqref{HA&B}. Hence, the global evolution of the two networks will lead to the
state
\begin{align}
& \frac{1}{\sqrt{2}}\left(  |0\rangle_{A}e^{\iota H(G_{B})t}|k\rangle
_{B}+e^{\iota H(G_{A})t}|i\rangle_{A}|0\rangle_{B}\right)  \nonumber\\
& =\frac{1}{\sqrt{2}}\left(  |0\rangle_{A}|\varphi_{k}\rangle_{B}+|\psi
_{i}\rangle_{A}|0\rangle_{B}\right)  ,
\end{align}
where $|\psi_{i}\rangle_{A}=e^{\iota H(G_{A})t}|i\rangle_{A}$ and
$|\varphi_{k}\rangle_{B}=e^{\iota H(G_{B})t}|k\rangle_{B}$.

To see how entanglement propagates from the sites $i$ and $k$, we evaluate the
concurrence \cite{wkw} between two generic sites $j$ and $l$ (different from
$i,k$). To this end, we have to consider the following density matrix
$\rho_{AB}$:%
\begin{eqnarray}
\rho_{AB}  & =&\frac{1}{2}\mathrm{Tr}_{\not j}\{|0\rangle_{A}\langle
0|\}\otimes\mathrm{Tr}_{\not \ell}\{|\varphi_{k}\rangle_{B}\langle\varphi
_{k}|\}\nonumber\label{trace2net}\\
& +&\frac{1}{2}\mathrm{Tr}_{\not j}\{|0\rangle_{A}\langle\psi_{i}%
\}|\otimes\mathrm{Tr}_{\not \ell}\{|\varphi_{k}\rangle_{B}\langle
0|\}\nonumber\\
& +&\frac{1}{2}\mathrm{Tr}_{\not j}\{|\psi_{i}\rangle_{A}\langle0|\}\otimes
\mathrm{Tr}_{\not \ell}\{|0\rangle_{B}\langle\varphi_{k}|\}\nonumber\\
& +&\frac{1}{2}\mathrm{Tr}_{\not j}|\psi_{i}\rangle_{A}\langle\psi_{i}%
|\otimes\mathrm{Tr}_{\not \ell}\{|0\rangle_{B}\langle0|\},\label{co}%
\end{eqnarray}
where $\mathrm{Tr}_{\not j}$ denotes the trace overall qubits but the $j$-th one.

For a single pure excitation of a spin-half network, the wave functions can be
expanded as
\begin{equation}
|\psi\rangle=\alpha_{1}|1\rangle+\ldots+\alpha_{n}|n\rangle,
\end{equation}
or
\begin{equation}
|\varphi\rangle=\beta_{1}|1\rangle+\ldots+\beta_{n}|n\rangle.
\end{equation}
In this way it is possible to express the density matrix $\rho_{AB}$ as a
function of the complex coefficients $\alpha$ and $\beta$:

\begin{center}
$\rho_{AB}=\frac{1}{2}\left(
\begin{array}
[c]{cccc}%
\sum_{i\neq j}^{n}|\alpha_{i}|^{2}+\sum_{k\neq\ell}^{n}|\beta_{k}|^{2} & 0 &
0 & 0\\
0 & |\beta_{\ell}|^{2} & \alpha_{j}^{*}\beta_{\ell} & 0\\
0 & \alpha_{j}\beta_{\ell}^{*} & |\alpha_{j}|^{2} & 0\\
0 & 0 & 0 & 0\\
&  &  &
\end{array}
\right) $.
\end{center}

For the single excitation model including both networks with Hamiltonians
\eqref{HA&B}, the $\rho_{AB}\check{\rho}_{AB}$ matrix is
\begin{align}
& \rho_{AB}\check{\rho}_{AB}\nonumber\label{rhorhotil}\\
& =\frac{1}{4}\left(
\begin{array}
[c]{cccc}%
0 & 0 & 0 & 0\\
0 & 2|\alpha_{j}|^{2}|\beta_{\ell}|^{2} & |\alpha_{j}|^{2}(\alpha_{j}^{\ast
}\beta_{\ell}+\alpha_{j}\beta_{\ell}^{\ast}) & 0\\
0 & |\beta_{\ell}|^{2}(\alpha_{j}^{\ast}\beta_{\ell}+\alpha_{j}\beta_{\ell
}^{\ast}) & 2|\alpha_{j}|^{2}|\beta_{\ell}|^{2} & 0\\
0 & 0 & 0 & 0\\
&  &  &
\end{array}
\right)  .\nonumber\\
&
\end{align}
Here, accordingly to \cite{wkw}
\begin{equation}
\check{\rho}_{AB}=(Y_{A}\otimes Y_{B})\rho_{AB}^{\ast}(Y_{A}\otimes
Y_{B}),
\end{equation}
where $\rho_{AB}^{\ast}$ is the complex conjugate of $\rho_{AB}$ \cite{wkw}.

The concurrence between two qubits $j$ and $l$ is defined as \cite{wkw}
\begin{equation}
C_{j,l}(\rho)=\max\left(0, {\sqrt{\lambda_{1}}-\sqrt{\lambda_{2}}-\sqrt
{\lambda_{3}}-\sqrt{\lambda_{4}}}\right)  ,
\end{equation}
where $\lambda_{1}$, $\lambda_{2}$, $\lambda_{3}$, and $\lambda_{4}$ are the
eigenvalues of $\rho_{AB}\check{\rho}_{AB}$ in the decreasing order.

Explicitly,
\begin{equation}
C_{j,l}=|\alpha_{j}||\beta_{\ell}|.
\end{equation}
It turns out that the concurrence is equal to the product of the fidelity reached
during the transmission of a single excitation in each network. As it is expected,
if there is perfect state transfer between particle $i$ and particle $j$ on $A$ as
well as between particles $k$ and $l$ on $B$, entanglement is perfectly
transferred from the pair $i,k$ to the pair $j,l$.


\section{Entanglement concentration}

We now consider two identical networks sharing pairs of qubits in a pure, non-
maximally entangled state and look for a possibility of enhancing the amount of
such entanglement.

\subsection{Two qubits networks}

In the simplest configuration we may consider two sites on each network
connected by an edge. These networks are simply two graphs consisting of a
single edge. Then, the resulting Hamiltonians $H_{A}$ and $H_{B}$ correspond
to the adjacency matrices
\begin{equation}
\mathbf{\emph{A}}(G_{A,B};P_{2})=\left(
\begin{array}
[c]{cc}%
0 & 1\\
1 & 0
\end{array}
\right) ,
\end{equation}
where $P_{n}$ denotes an open chain of length $n-1$.
Let the initial state of qubits $A_{1}$ and $B_{1}$ be%
\begin{equation}
|\psi\rangle=\cos(\theta)|0\rangle_{A_{1}}|0\rangle_{B_{1}}+\sin
(\theta)|1\rangle_{A_{1}}|1\rangle_{B_{1}}\label{psi0}%
\end{equation}
and the initial state of qubits $A_{2}$ and $B_{2}$ be $|0\rangle_{A_{2}%
}|0\rangle_{B_{2}}$. The two networks will evolve independently, so that the
whole state
\begin{align}
|\Psi\rangle & =\cos(\theta)|0\rangle_{A_{1}}|0\rangle_{B_{1}}|0\rangle
_{A_{2}}|0\rangle_{B_{2}}\label{psi}\\
& +\sin(\theta)|1\rangle_{A_{1}}|1\rangle_{B_{1}}|0\rangle_{A_{2}}%
|0\rangle_{B_{2}}\nonumber
\end{align}
will have the following general form at time \emph{t}:%
\begin{align}
|\Psi(t)\rangle & =\cos(\theta)|0\rangle_{A_{1}}|0\rangle_{A_{2}}%
|0\rangle_{B_{1}}|0\rangle_{B_{2}}\nonumber\label{psi}\\
& +\sin(\theta)\cos^{2}(t)|1\rangle_{A_{1}}|0\rangle_{A_{2}}|1\rangle_{B_{1}%
}|0\rangle_{B_{2}}\nonumber\\
& +\sin(\theta)\cos(t)\sin(t)|1\rangle_{A_{1}}|0\rangle_{A_{2}}|0\rangle
_{B_{1}}|1\rangle_{B_{2}}\nonumber\\
& +\sin(\theta)\cos(t)\sin(t)|0\rangle_{A_{1}}|1\rangle_{A_{2}}|1\rangle
_{B_{1}}|0\rangle_{B_{2}}\nonumber\\
& +\sin(\theta)\sin^{2}(t)|0\rangle_{A_{1}}|1\rangle_{A_{2}}|0\rangle_{B_{1}%
}|1\rangle_{B_{2}}.
\end{align}

Actually, given the evolution, also the qubits $A_{2}$ and $B_{2}$ become
entangled with $A_{1}$ and $B_{1}$. Local measurements on $A_{2}$ and $B_{2}$
may concentrate the entanglement available at time $t$ on qubits $A_{1}$ and
$B_{1}$, possibly increasing the initial amount. Suppose
that we locally measure $Z_{A_{2}}\otimes Z_{B_{2}}$. For outcomes $00$, the
state in Eq. \eqref{psi} will be projected onto $|0\rangle_{A_{2}}%
|0\rangle_{B_{2}}$. Hence, the state $\sigma=|\Psi(t)\rangle\langle\Psi(t)|$
changes into
\begin{equation}
\sigma\mapsto\rho_{A_{1}B_{1}}\otimes|0\rangle_{A_{2}}\langle0|\otimes
|0\rangle_{B_{2}}\langle0|,\label{sig}%
\end{equation}
where
\begin{equation}
\rho_{A_{1}B_{1}}=\frac{\tilde{\rho}_{A_{1}B_{1}}}{Tr_{A_{1}B_{1}}%
\{\tilde{\rho}_{A_{1}B_{1}}\}},\label{rho}%
\end{equation}
and
\begin{equation}
\tilde{\rho}_{A_{1}B_{1}}=_{A_{2}}\langle0|_{B_{2}}\langle0|\sigma
|0\rangle_{B_{2}}|0\rangle_{A_{2}}.\label{rho'}%
\end{equation}
It results
\begin{align}
\tilde{\rho}_{A_{1}B_{1}}  & =_{A_{2}}\langle0|_{B_{2}}\langle0||\Psi
(t)\rangle\langle\Psi(t)||0\rangle_{B_{2}}|0\rangle_{A_{2}}\label{rhot}\\
& =(\cos(\theta)|0\rangle_{A_{1}}|0\rangle_{B_{1}}+\sin(\theta)\cos
^{2}(t)|1\rangle_{A_{1}}|1\rangle_{B_{1}})\nonumber\\
& (\cos(\theta)_{A_{1}}\langle0|_{B_{1}}\langle0|+\sin(\theta)\cos
^{2}(t)_{A_{1}}\langle1|_{B_{1}}\langle1|).\nonumber
\end{align}



The probability of this transformation is
\begin{equation}
P_{00}(t)=\mathrm{Tr}_{A_{1}B_{1}}\{\tilde{\rho}_{A_{1}B_{1}}\}=\cos^{2}(\theta)+\sin^{2}(\theta)\cos^{4}(t).\label{prob}%
\end{equation}

Notice that the state \eqref{rho} with \eqref{rho'} and \eqref{prob} is quite generally
entangled with a concurrence possibly higher then the initial state \eqref{psi}. On
the contrary, it is easy to check that the state resulting from the projection of
Eq.\eqref{psi} onto $|0\rangle_{A_{2}}|1\rangle_{B_{2}}$, or
$|1\rangle_{A_{2}}|0\rangle_{B_{2}}$, or $|1\rangle_{A_{2}}|1\rangle_{B_{2}}$
(corresponding to outcomes $01$, $10$, $11$ respectively) is separable, hence
with zero concurrence.

Since the enhancement of the initial entanglement is determined by the
measurement outcomes, we have to account for their respective probabilities in
order to evaluate the performance of the protocol. Then, we define the
efficiency of the protocol as the average of the increments of entanglement
weighted by the respective probabilities, that is,
\begin{equation}
E:=\sum_{o}P_{o}\,\Delta_{o}C.\label{Edef}%
\end{equation}
Here the index $o$ runs over all possible measurement outcomes, each
occurring with probability $P_{o}$. Moreover
\begin{equation}
\Delta_{o}C:=\left\{
\begin{tabular}
[c]{ll}%
$(C_{o}-C),$ & if $C_{o}>C$;\\
$0,$ & if $C_{o}\leq C$.
\end{tabular}
\ \right.
\end{equation}
Here $C$ is the concurrence of $\rho_{A_{1}B_{1}}$ and $C_{o}$ the concurrence
of $\rho_{A_{1}B_{1}}^{\prime}$ conditioned to the outcome $o$.

In our case the concurrence for the outcome $00$ reads
\begin{equation}\label{Coo}
    C_{00}=\frac{2\sin(\theta)\cos(\theta)\cos^{2}(t)}{\cos^{2}(\theta)+\sin^{2}(\theta)\cos^{4}(t)},
\end{equation}

\pagebreak

thus the efficiency simply results
\begin{eqnarray}\label{eff1pair}
&&E(\theta,t)=P_{00}\Delta_{00}C\\
&&=2\sin(\theta)\cos(\theta)(\cos^{2}(t)-\cos^{2}(\theta)-\sin^{2}(\theta)\cos^{4}(t)).
\nonumber
\end{eqnarray}
Now taking $\frac{\partial E (\theta,t)}{\partial t}=0$ we get the optimal time (maximizing \eqref{eff1pair}) as
\begin{equation}\label{opt}
\cos^{2}(t)=\frac{1}{2\sin^{2}(\theta)}.
\end{equation}
By substituting it back to Eq.(\ref{eff1pair}) we arrive at the maximum efficiency
written as solely function of the variable $\theta$
\begin{equation}\label{eff1pair2}
    E(\theta)=\frac{\cos(\theta)}{2\sin(\theta)}(1-4\cos^{2}(\theta)\sin^{2}(\theta)).
\end{equation}
The behavior of this quantity is depicted in Fig. (\ref{concentration}).


Let us now consider the case of two initially entangled pairs. Let Eq. \eqref{psi0}
be the initial state of qubits $A_{1}$ and $B_{1}$ as well as qubits $A_{2}$ and
$B_{2}$. Write $v_{A_{i}B_{i}}=|\psi\rangle\langle\psi|$, then the initial state of the
networks simply results
\begin{equation}
v_{A_{1}B_{1}}\otimes v_{A_{2}B_{2}},\label{ini2}%
\end{equation}
and it evolves into
\begin{equation}
\sigma=e^{-\iota H_{A}t}e^{-\iota H_{B}t}\left(  v_{A_{1}B_{1}}\otimes
v_{A_{2}B_{2}}\right)  e^{\iota H_{B}t}e^{\iota H_{A}t}.
\end{equation}

Now let us consider local measurements of the Pauli observables $Z_{A_{2}%
}\otimes Z_{B_{2}}$, \emph{i.e.}, only one out of the two initial pairs is locally
measured.

For outcome $a_{2}b_{2}$ (with $a_{2},b_{2}\in\{0,1\}$) the above state will
be projected onto $|a_{2}\rangle_{A_{2}}|b_{2}\rangle_{B_{2}}$. Hence,
\begin{equation}
\sigma\mapsto\rho_{A_{1}B_{1}}\otimes|a_{2}\rangle_{A_{2}}\langle
a_{2}|\otimes|b_{2}\rangle_{B_{2}}\langle b_{2}|,
\end{equation}
where
\begin{equation}
\rho_{A_{1}B_{1}}=\frac{\tilde{\rho}_{A_{1}B_{1}}}{\mathrm{Tr}_{A_{1}B_{1}%
}\{\tilde{\rho}_{A_{1}B_{1}}\}},
\end{equation}
and
\begin{equation}
\tilde{\rho}_{A_{1}B_{1}}=\langle a_{2}b_{2}|\sigma|b_{2}a_{2}\rangle.
\end{equation}

Outcome $a_{2}b_{2}=00$ occurs at time $t$ with probability
\begin{eqnarray}\label{prob01}
P_{00}(t)  &=&\mathrm{Tr}_{A_{1}B_{1}}\{\tilde{\rho}_{A_{1}B_{1}}\}\\
\nonumber&=&\frac{1}{2}\cos^{2}(\theta)(2\cos^{2}(\theta)-\sin^{2}(\theta)-\sin^{2}(\theta)\cos(4t)),
\end{eqnarray}
and the corresponding conditional state results (in matrix form)
\begin{eqnarray}
&&\tilde \rho_{A_{1}B_{1}}(t)=\\
&&\left(
\begin{array}
[c]{cccc}%
\cos^{4}(\theta) & 0 &0 &\cos^{3}(\theta)\sin(\theta)\cos(2t)\\
0 & 0 & 0 & 0\\
0 & 0 &0 & 0\\
\cos^{3}(\theta)\sin(\theta)\cos(2t) & 0 & 0 &\cos^{2}(\theta)\sin^{2}(\theta)\cos^{2}(2t)\\
&  &  &
\end{array}
\right).\nonumber
\end{eqnarray}

Its concurrence reads
\begin{equation}\label{C012}
    C_{00}(t)=\frac{4\sin(\theta)\cos(\theta)\cos(2t)}{2\cos^{2}(\theta)-\sin^{2}(\theta)-\sin^{2}(\theta)\cos(4t)}.
\end{equation}

 Outcomes $a_{2}b_{2}=11$ occurs at time $t$ with probability
\begin{eqnarray}\label{prob10}
P_{11}(t)  & =&\mathrm{Tr}_{A_{1}B_{1}}\{\tilde{\rho}_{A_{1}B_{1}}\}\\
\nonumber&=&\frac{1}{2}\sin^{2}(\theta)(\cos^{2}(\theta)+-2\sin^{2}(\theta)+\cos^{2}(\theta)\cos(4t)),
\end{eqnarray}
and the corresponding conditional state results (in matrix form)
\begin{eqnarray}
&&\tilde\rho_{A_{1}B_{1}}(t)=\\
&&\left(
\begin{array}
[c]{cccc}%
cos^{2}(\theta)\sin^{2}(\theta)\cos^{2}(2t) & 0 &0 &\cos(\theta)^{3}\sin(\theta)\cos(2t)\\
0 &0 & 0& 0\\
0 & 0 &0 & 0\\
\cos(\theta)\sin^{3}(\theta)\cos(2t) & 0 & 0 &\sin^{4}(\theta)\\
&  &  &
\end{array}
\right).\nonumber
\end{eqnarray}
Its concurrence reads
\begin{equation}\label{C012}
    C_{11}(t)=\frac{4\sin(\theta)\cos(\theta)\cos(2t)}{\cos^{2}(\theta)-2\sin^{2}(\theta)+\cos^{2}(\theta)\cos(4t)}.
\end{equation}

Since the other outcomes ($a_{2}b_{2}=01,10$) give separable conditional
states, the efficiency in Eq.\eqref{Edef} becomes
\begin{eqnarray}\label{eff2pair}
E(\theta,t) &=&\cos^{4}(\theta)\sin^{4}(\theta)\nonumber\\
&\times&(\cos^{2}(\theta)-2\sin^{2}(\theta)-2\cos(2t)+\cos^{2}(\theta)\cos(4t))\nonumber\\
&\times&(2\cos^{2}(\theta)-\sin^{2}(\theta)-2\cos(2t)-\sin^{2}(\theta)\cos(4t))\nonumber\\
&-&2\cos^{5}(\theta)\sin(\theta)\cos(4t).
\end{eqnarray}
Now, by taking $\frac{\partial E (\theta,t)}{\partial t}=0$ we get the optimal time (maximizing \eqref{eff2pair}) as
$t=\pi$. By substituting it back to Eq. (\ref{eff2pair}) we arrive at the maximum
efficiency written as solely function of the variable $\theta$
\begin{eqnarray}
E(\theta) &=&2\cos^{4}(\theta)\sin(\theta)(8\sin^{7}(\theta)-\cos(\theta))\nonumber\\
\end{eqnarray}
The behavior of this quantity is depicted in Fig. (\ref{concentration}) for $\pi/4\leq\theta\leq \pi/2$.


\subsection{Networks with more than two qubits}

\begin{figure}[ptb]
\begin{tabular}{|c|}
  \hline
  \includegraphics[width=250 pt]{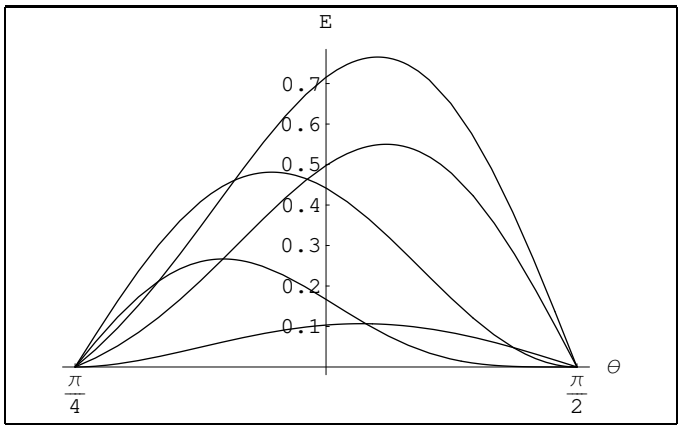}\\
  \hline
\end{tabular}
\caption{Maximum efficiency
$E(\theta)$ as a function of $\theta$ for different networks.Curves from bottom up
refer to networks with number of nodes and number of entangled pairs equal to
$(2,1)$, $(2,2)$, $(3,3)$, $(4,4)$, $(5,5)$ respectively.}
\label{concentration}%
\end{figure}

In the previous subsection we have seen that having initially a single entangled
pure pair suffices to achieve entanglement concentration. However, looking at
Fig. (\ref{concentration}), we can conclude that having initially two entangled pure
pairs give to the protocol a better performance. From now on we are going to
consider a number of initially entangled pure pairs equal to the number of vertices
in the network. For $n$ vertices, consider the initial state of qubits $A_{i}$ and
$B_{i}$ to be \eqref{psi0}. Writing $v_{A_{i}B_{i}}=|\psi_{i}\rangle\langle\psi_{i}|$,
the initial state of the system is then
\begin{equation}
V_{12\ldots n}=v_{A_{1}B_{1}}\otimes v_{A_{2}B_{2}}\otimes\ldots\otimes
v_{A_{n}B_{n}}.\label{ini3}%
\end{equation}
The time-evolution gives
\begin{equation}
\sigma=e^{-\iota H_{A}t}e^{-\iota H_{B}t}V_{12\ldots n}e^{\iota H_{B}%
t}e^{\iota H_{A}t}.
\end{equation}
Now we measure with respect to the Pauli observables $(Z_{A_{2}}\otimes
\ldots\otimes Z_{A_{n}})\otimes(Z_{B_{2}}\otimes\ldots\otimes Z_{B_{n}})$,
\emph{i.e.}, $n-1$ out of initially $n$ pairs are locally subjected to measurement.

When we obtain $a_{2}\ldots a_{n}b_{2}\ldots b_{n}$, where each variable is
either zero or one, the above state will be projected onto
$|a_{2}\rangle_{A_{2}}\ldots|a_{n}\rangle_{A_{n}}|b_{2}\rangle_{B_{2}}%
\ldots|b_{n}\rangle_{B_{n}}$. Then
\begin{align}
\sigma\mapsto\rho_{A_{1}B_{1}}  & \otimes|a_{2}\rangle_{A_{2}}\langle
a_{2}|\otimes\ldots\otimes|a_{n}\rangle_{A_{n}}\langle a_{n}|\nonumber\\
& \otimes|b_{2}\rangle_{B_{2}}\langle b_{2}|\otimes\ldots\otimes|b_{n}%
\rangle_{B_{n}}\langle b_{n}|,
\end{align}
where
\begin{equation}
\rho_{A_{1}B_{1}}=\frac{\tilde{\rho}_{A_{1}B_{1}}}{\mathrm{Tr}_{A_{1}B_{1}%
}\{\tilde{\rho}_{A_{1}B_{1}}\}},
\end{equation}
and
\begin{equation}
\tilde{\rho}_{A_{1}B_{1}}=\langle a_{2}\ldots a_{n}b_{2}\ldots b_{n}%
|\sigma|b_{n}\ldots b_{2}a_{n}\ldots a_{2}\rangle.
\end{equation}
By taking into account all $4^{n-1}$ possible outcomes we can then calculate
the efficiency given in Eq. \eqref{Edef}.

Let us now consider $n=3$ and the two networks in an open chain configuration.
Then, the Hamiltonians $H_{A}$ and $H_{B}$ come from the following adjacency
matrices
\begin{equation}
\mathbf{\emph{A}}(G_{A,B};P_{3})=\left(
\begin{array}
[c]{ccc}%
0 & 1 & 0\\
1 & 0 & 1\\
0 & 1 & 0
\end{array}
\right)  .\label{adj31}%
\end{equation}
The efficiency is maximized by $t=k\pi/\sqrt{2}$, $k\in\mathbb{N}$ giving
\begin{align}
E(\theta)=5\cos^{2}(\theta)\sin^{2}(\theta)(1-2\cos^{2}(\theta)),
\label{Efficiency}
\end{align}
where $\pi/4\leq\theta\leq\pi/2$.

The same efficiency can be obtained by considering the dynamics resulting from the
fully connected three qubits networks. The related Hamiltonians $H_{A}$ and
$H_{B}$ are
\begin{equation}
\mathbf{\emph{A}}(G_{A,B};K_{3})=\left(
\begin{array}
[c]{ccc}%
0 & 1 & 1\\
1 & 0 & 1\\
1 & 1 & 0
\end{array}
\right),\label{adj32}%
\end{equation}
where $K_{n}$ is the fully connected graph with $n$ vertices. However, in this
case, the expression (\ref{Efficiency}) is obtained at time $t=k\pi$,
$k\in\mathbb{N}$. The quantity (\ref{Efficiency}) is represented in Fig.
(\ref{concentration}).


Practically, the change of the underlying graph associated to the network only
affects the time of the dynamics, but it does not affect the expression of the
maximum efficiency. We conjecture (supported by numerical evidence) that this
is also true for networks having more than three nodes, provided that there are no
isolated nodes. In fact, the protocol involves entangled pairs shared by single
nodes of each network. Since we consider a number of initially entangled pairs
equal to the number of nodes in a network, two isolated nodes that share an
entangled pair would constitute a separate system. Then, the maximum
efficiency only depends on the number of nodes and so on the total number of
initially entangled pairs.


Let us now consider $n=4$ and suppose that the two networks $A$ and $B$ are
both in a closed chain configuration (a cycle).
Then, the Hamiltonians $H_{A}$ and
$H_{B}$ result from the following adjacency matrices
\begin{equation}
\mathbf{\emph{A}}(G_{A,B};C_{4})=\left(
\begin{array}
[c]{cccc}%
0 & 1 & 0 & 1\\
1 & 0 & 1 & 0\\
0 & 1 & 0 & 1\\
1 & 0 & 1 & 0
\end{array}
\right),\label{adj42}%
\end{equation}
where $C_{n}$ is the closed chain of length $n-1$. In this case the maximum
efficiency is obtained for $t=k\pi$, $k\in\mathbb{N}$ and it reads
\begin{equation}
E(\theta)=\frac{1}{16}\cos(\theta)\sin^{3}(\theta)(8\cos(4\theta)-3\cos(6\theta)-29\cos(2\theta)+8),\label{Efficiency4}%
\end{equation}
where $\pi/4\leq\theta\leq\pi/2$. Such efficiency is attained at $t=\pi/2+k\pi$,
$k\in\mathbb{N}$, in the case of fully connected configurations.

The efficiency (\ref{Efficiency4}) is represented in Fig. (\ref{concentration}).

Let us finally consider the case of $n=5$.
We chose Hamiltonians
$H_{A}$ and $H_{B}$ coming from the following adjacency matrices
\begin{equation}
\mathbf{\emph{A}}(G_{A,B};C_{5})=\left(
\begin{array}
[c]{ccccc}
0 & 1 & 0 & 0 & 1\\
1 & 0 & 1 & 0 & 0\\
0 & 1 & 0 & 1 & 0\\
0 & 0 & 1 & 0 & 1\\
1 & 0 & 0 & 1 & 0
\end{array}
\right).\label{adj41}%
\end{equation}
The maximum efficiency results
\begin{equation}
E(\theta)=\frac{1}{4}\cos(\theta)\sin^{3}(\theta)(\cos(4\theta)-\cos(6\theta)-13\cos(2\theta)+1),\label{Efficiency5}%
\end{equation}
where $\pi/4\leq\theta\leq\pi/2$. This is obtained for $t=k\pi$, $k\in\mathbb{N}$.
The efficiency (\ref{Efficiency5}) is represented in Fig.\ref{concentration}.

\section{Entanglement Purification}

We shall consider pairs of qubits in networks $A$ and $B$ initially in a mixed
entangled state. We then determine conditions for enhancing such entanglement.
Specifically, we consider the initial state of qubits $A_{i}B_{i}$ to be the
Werner mixture \cite{werner89} written as
\begin{align}
& w_{A_{i}B_{i}}=f\,|\Phi^{+}\rangle_{A_{i}B_{i}}\langle\Phi^{+}|+\frac
{1-f}{3}\,\left(  |\Phi^{-}\rangle_{A_{i}B_{i}}\langle\Phi^{-}|\right.
\nonumber\\
& \hspace{1cm}\left.  +|\Psi^{+}\rangle_{A_{i}B_{i}}\langle\Psi^{+}|+|\Psi
^{-}\rangle_{A_{i}B_{i}}\langle\Psi^{-}|\right)  ,\label{Werner}%
\end{align}
where
\begin{align}
|\Phi^{\pm}\rangle_{A_{i}B_{i}}  & =\frac{|0\rangle_{A_{i}}|0\rangle_{B_{i}%
}\pm|1\rangle_{A_{i}}|1\rangle_{B_{i}}}{\sqrt{2}},\\
|\Psi^{\pm}\rangle_{A_{i}B_{i}}  & =\frac{|0\rangle_{A_{i}}|1\rangle_{B_{i}%
}\pm|1\rangle_{A_{i}}|0\rangle_{B_{i}}}{\sqrt{2}},
\end{align}
are the Bell states and $f$ is the fidelity between $w$ and $|\Phi^{+}\rangle$
(notice that $0.25\le f\le 1$). The initial state of the networks is
\begin{equation}
W_{12\ldots n}=w_{A_{1}B_{1}}\otimes w_{A_{2}B_{2}}\otimes\ldots\otimes
w_{A_{n}B_{n}}.\label{ini4}%
\end{equation}
By using the same procedure shown for entanglement concentration, we can
calculate the efficiency of the protocol. Networks with one or two qubits lead to a
zero efficiency. In fact, the concurrence obtained after the evolution of the initial
state in Eq. (\ref{psi}) is less than the initial one and it is identical for every
outcomes. Actually for all outcomes $o$ we have
\begin{equation}
C_{o}-C=(1+4f-32f^{2})/36\leqslant0.\label{twopairs}%
\end{equation}

\begin{figure}[ptb]
\begin{tabular}{|c|}
  \hline
  \includegraphics[width=250 pt]{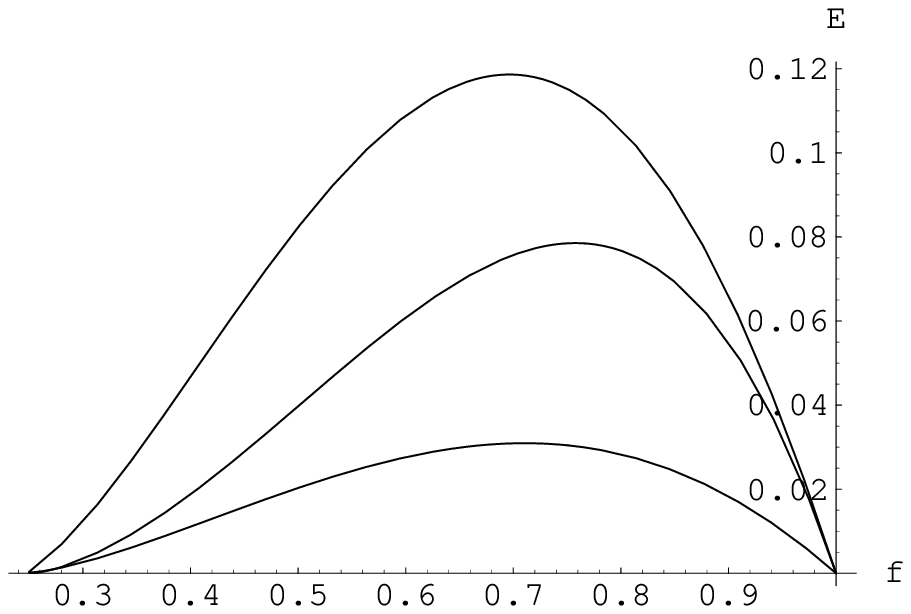}\\
  \hline
\end{tabular}
\caption{Maximum efficiency
$E(f)$ as a function of $f$ for different networks. Curves from bottom up refer
to networks with number of nodes and number of entangled pairs equal to
$(3,3)$, $(4,4)$, $(5,5)$ respectively. }%
\label{purification}%
\end{figure}

The efficiency for three qubits becomes greater than zero and its maximum is represented in Fig. (\ref{purification}).
Such maximum is obtained when $t=k\pi/\sqrt{2}$,
$k\in\mathbb{N}$, in the open chain configuration (from \eqref{adj31} and when $t=k\pi$,
$k\in\mathbb{N}$, in the fully connected configuration (from \eqref{adj32}).

Likewise the case of entanglement concentration,
only the time $t$ at which maximum efficiency is attained
is affected by the choice of the networks' topology (not the maximum efficiency itself), provided that there are no isolated nodes.

The maximum efficiency for $n=4$ and $n=5$ can be obtained from Eqs.
\eqref{adj42} and \eqref{adj41}, respectively. Their behavior is represented in Fig.
(\ref{purification}).

\section{Concluding Remarks}

In conclusion, we have considered entanglement manipulation between two
networks of spin-half particles with $XY$-type interaction. We have related
entanglement transmission to the fidelity achieved in state transfer when
adopting the single excitation model.

We have shown how to increase the entanglement of pure and mixed states via
free dynamics and local measurements on the nodes. The performance of the
protocol are characterized by means of an efficiency measure. The performance
appears to be independent of the networks' topology provided that there are no
isolated nodes. The structure of the network only affects the time at which the
maximum efficiency is attained. Moreover, we have determined the minimal
number of initial entangled pairs shared between the two networks and needed to
successfully complete the protocol (\emph{i.e.}, with nonzero efficiency). This
number is 1 and 3 for entanglement concentration and purification, respectively.
Furthermore, the results shown in Figs. (\ref{concentration}) and
(\ref{purification}) suggest that a (quasi) constant increase for the maxima of the
efficiency is obtained with a corresponding increase for the number of entangled
pairs. This can be estimated as about $0.15$, for the case of entanglement
concentration, and about $0.04$, for the case of purification for each additional
pair. This trend suggests that a (quasi) deterministic protocol can be achieved, at
least for specific values of $\theta$ or $f$, with a limited number of initially
entangled pairs. Our protocol is straightforward and only based on the free
dynamics occurring in the two networks. It may be an interesting venue to
consider more general settings, as, for example, synchronous dynamics in
different networks.

Notice that already in the single excitation model the XYZ (Heisenberg)
interaction would give a different networks' dynamics. Actually a phase factor
would be introduced in front of the state's components containing one excitation
per network. However such a phase factor is not influent to our protocol as long
as we consider measurements on the computational basis (Z eigenvectors).
Thus our conclusions can be applied to XY model as well as to Heisenberg
model.

Finally, it is worth remarking that the proposed schemes for entanglement
manipulation might be experimentally verified with networks containing few
spins-half systems. Candidate physical systems would be semiconductor quantum dots
\cite{loss98}, superconducting qubits \cite{you05} or atoms trapped in optical
lattices \cite{mandel03}.


\acknowledgments While writing this paper, AC would like to thank Seth Lloyd
for the kind of hospitality at Research Laboratory of Electronics (MIT) where part
of this work was carried out. The work of SM is supported by the European
Commission, under the FET-Open grant agreement HIP, number
FP7-ICT-221889. SS is supported by a Newton International Fellowship. The
authors would like to thank Yasser Omar for useful discussions.


\end{document}